\documentclass[preprint2]{aastex}

\shorttitle{UCAC3 Proper Motion Survey II}
\shortauthors{Finch}

\begin{document}

\title{UCAC3 Proper Motion Survey. II. \\ DISCOVERY OF NEW PROPER
  MOTION STARS IN UCAC3 \\ WITH 0$\farcs$40 yr$^{-1}$ $>$ $\mu$ $\ge$
  0$\farcs$18 yr$^{-1}$ BETWEEN DECLINATIONS $-$47$\degr$ and
  00$\degr$}

\author{Charlie T. Finch, Norbert Zacharias}

\email{finch@usno.navy.mil}

\affil{U.S. Naval Observatory, Washington DC 20392--5420}

\author{Mark R. Boyd, Todd J. Henry}

\affil{Georgia State University, Atlanta, GA 30302--4106}

\author{Nigel C. Hambly}

\affil{Scottish Universities Physics Alliance, Institute for
Astronomy, University of Edinburgh, Royal Observatory, Blackford Hill,
Edinburgh EH9 3HJ, Scotland, UK}


\begin{abstract}

We present 474 new proper motion stellar systems in the southern sky
having no previously known components, with 0$\farcs$40 yr$^{-1}$ $>$
$\mu$ $\ge$ 0$\farcs$18 yr$^{-1}$ between declinations $-$47$\degr$
and 00$\degr$.  In this second paper utilizing the U.S. Naval
Observatory third CCD Astrograph Catalog (UCAC3) we complete our sweep
of the southern sky for objects in the proper motion range targeted by
this survey with R magnitudes ranging from 9.80 to 19.61.  The new
systems contribute a $\sim$16\% increase in the number of new stellar
systems for the same region of sky reported in previous SuperCOSMOS
RECONS (SCR) surveys.  Among the newly discovered stellar systems are
16 multiples, plus an additional 10 components that are new common
proper motion companions to previously known objects.  A comparison of
UCAC3 proper motions to those from Hipparcos, Tycho-2, Southern Proper
Motion (SPM4), and SuperCOSMOS indicates that all proper motions are
consistent to $\sim$10 mas/yr, with the exception of SuperCOSMOS.
Distance estimates are derived for all stellar systems having
SuperCOSMOS Sky Survey (SSS) $B_J$, $R_{59F}$, and $I_{IVN}$ plate
magnitudes and Two-Micron All Sky Survey (2MASS) infrared photometry.
We find five new red dwarf systems estimated to be within 25 pc.
These discoveries support results from previous proper motion surveys
suggesting that more nearby stellar systems are to be found,
particularly in the fainter, slower moving samples.

In this second paper utilizing the U.S. Naval Observatory third CCD
Astrograph Catalog (UCAC3) we complete our sweep of the southern sky
for objects in the proper motion range targeted by this survey with R
magnitudes ranging from 9.80 to 19.61.

\end{abstract}


\section{INTRODUCTION}

The third U.S. Naval Observatory (USNO) CCD Astrograph Catalog
(UCAC3)\\ \citep{u3r} proper motion survey, addresses the possibility
that proper motion surveys using digitized scans of photographic
plates might overlook some proper motion systems.  The UCAC3 obtained
accurate proper motions by combining CCD observations with early epoch
photographic data.  This survey utilizes the UCAC3 proper motions to
discover new systems that have been missed in previous efforts.  The
first paper in this series \citep{upm1} (hereafter, U3PM1), confirmed
this suspicion by revealing an additional 25.3\% stellar systems
having a proper motion of 0$\farcs$40 yr$^{-1}$ $>$ $\mu$ $\ge$
0$\farcs$18 yr$^{-1}$ between declinations $-$90$\degr$ and
$-$47$\degr$ over those found by the Research Consortium On Nearby
Stars (RECONS)\footnote{\it www.recons.org} group using SuperCOSMOS
Sky Survey (SSS) data.  These new discoveries provided the impetus for
this second paper of the series, which completes the sweep of the
southern sky for systems with 0$\farcs$40 yr$^{-1}$ $>$ $\mu$ $\ge$
0$\farcs$18 yr$^{-1}$ found in the UCAC3.

The data obtained from proper motion surveys aid astronomers in
determining accurate stellar luminosity and mass functions, thereby
revealing how the Galaxy's stellar mass is divided among different
types of stars.  Our main goal --- identifying the Sun's nearest
neighbors --- provides a vast sample of red dwarf, subdwarf, and white
dwarf stellar systems for studies of multiplicity, activity, ages, and
exoplanet searches.  Because of their proximity, the nearby stars
offer the most accessible measurements of each of these
characteristics.

Our UCAC3 proper motion survey is currently focused on the southern
hemisphere, which has not been surveyed as systematically as the
northern sky, where the pioneering surveys of Giclas
\citep{1971lpms.book.....G,1978LowOB...8...89G} and Luyten
\citep{1979lccs.book.....L,1980PMMin..55....1L} were primarily carried
out.  Historically, proper motion studies have been focused on
blinking photographic plates taken at different epochs to determine
source motions.  Recent surveys that complement the classic efforts
utilize various techniques, plate sets, modern computers, and
carefully tailored algorithms to effectively blink digitized images of
photographic plates.  In the southern sky, such surveys include
\citep{1994A&AS..105..179W}, \citep{1999A&AS..139...25W},
\citep{2000A&A...353..958S, 2002ApJ...565..539S},
\citep{2001Sci...292..698O}, \citep{2003A&A...397..575P},
\citep{2005AJ....130.1247L,2008AJ....135.2177L}, \citep{SIPS1,SIPS2},
and \citep{2009MNRAS.397.1685D}.

In an effort to understand the stellar population of the solar
neighborhood, the RECONS group has been targeting the neglected
southern sky to reveal new stellar proper motion systems.  To date,
these discoveries have been reported in six papers in {\it The Solar
Neighborhood} (TSN) series \citep{2004AJ....128..437H},
\citep{2004AJ....128.2460H}, \citep{2005AJ....129..413S},
\citep{2005AJ....130.1658S}, \citep{2007AJ....133.2898F},
\citep{tsnxxv}.  These new systems are discovered using the
SuperCOSMOS Sky Survey (SSS) data \citep{superc1} and given the name
SCR (SuperCOSMOS-RECONS).  Followup observations of intriguing systems
are performed at the Cerro Tololo Inter-American Observatory (CTIO)
0.9m telescope, where RECONS operates a trigonometric parallax program
focusing on stars within 25 pc.

Our UCAC3 survey uses an approach fundamentally different from plate
blinking to reveal proper motion systems.  We take advantage of
observations reported in many catalogs ranging in epochs from the
early nineteenth to the early twenty-first centuries, rather than
directly using specific sets of digitized images from photographic
plates.  In this investigation we focus on stellar systems in the
UCAC3 found between declinations $-$47$\degr$ and 00$\degr$ that have
0$\farcs$40 yr$^{-1}$ $>$ $\mu$ $\ge$ 0$\farcs$18 yr$^{-1}$,
completing a sweep of the southern sky.  The search region and proper
motion range matches that in \citep{tsnxxv}, hereafter TSN25, in which
the lower proper motion cutoff was chosen to match that of the NLTT
catalog.  TSN25 reports 2817 new SCR systems, substantially adding to
the number of new SCR systems found previously.  In
Table~\ref{diststat}, we summarize the number of new stellar systems
discovered, highlighting those estimated to be within 25 pc, for both
the RECONS and UCAC3 surveys.  In this paper we will focus in
particular on the two SCR searches (TSN18 and TSN25) that correspond
to the same proper motion and declination ranges as this UCAC3 survey
(U3PM1 and this paper).  New stellar objects from this search are
given USNO Proper Motion (UPM) names.

\section {Method}
\subsection {UCAC3}

The USNO CCD Astrograph Catalog (UCAC) project finished observations
in late 2004 and has been producing astrometric catalogs since October
2000.  This astrometric survey was conceived to densify the optical
reference frame to high accuracy beyond the Hipparcos and Tycho
magnitudes.  UCAC is the first all-sky survey performed with a CCD
detector utilizing the high level of precision achievable with this
technology.  The first release, UCAC1, \citep{2000AJ....120.2131Z},
was a partial catalog covering 80\% of the southern sky.  The second
catalog, UCAC2, \citep{2004AJ....127.3043Z}, contains roughly 80\% of
the entire sky and includes improved proper motions from the use of
early epoch plates paired with the Astrograph CCD data.  UCAC3
\citep{u3r}, released in August 2009, is the first in the series to
contain coverage of the entire sky.  UCAC3 also includes double star
fitting and has a slightly deeper limiting magnitude than UCAC2 due to
a complete re-reduction of the pixel data \citep{u3x}.  In addition,
data from the Two-Micron All Sky Survey (2MASS) were used in UCAC3 to
probe for and reduce systematic errors in UCAC observations, providing
a greater number of reference stars to stack up residuals as a
function of many parameters, such as observing site and exposure time.
A detailed description of the astrometric reductions of UCAC3 can be
found in \citep{u3a}.  A detailed introduction to the UCAC3 can be
found in the release paper \citep{u3r} and the README file of the data
distribution.  A new edition, UCAC4, \citep{u4r} is scheduled to be
released later this year.

\subsection {PROPER MOTIONS}

The UCAC3 contains roughly 95 million calculated absolute proper
motions.  The majority of these are derived proper motions from the
use of early epoch catalogs paired with the Astrograph CCD data.
Earlier epoch data are all reduced to the International Celestial
Reference Frame (ICRF).  UCAC3 mean positions and proper motions are
calculated using a weighted, least-squares adjustment procedure.

Bright stars with R$\sim$8--12 in UCAC3 are combined with ground-based
photographic and transit circle catalogs.  These include all catalogs
used for the production of the Tycho-2 project
\citep{2000A&A...355L..27H}, unpublished measures of over 5000
astrograph plates digitized on the StarScan machine \citep{starscan},
new reductions of Southern Proper Motion (SPM) \citep{spm} data, and
data from the SuperCOSMOS project \citep{superc1}.  About 1.2 million
star positions to about B$= $ 12 entered UCAC from digitizing the AGK2
plates (epoch about 1930).  The Hamburg Zone Astrograph and USNO Black
Birch Astrographs contributed another 7.3 million star positions,
mainly in the V$= $12 -- 14 magnitude range, in fields covering about
30\% of the sky, and the Lick Astrograph plates taken around 1990
yielded over one million star positions to V$= $16 in selected fields.

For all catalogs used to derive UCAC3 proper motions a systematic
error estimate was added to the root mean square (RMS) of the individual
stars random errors.  The largest error floor added was 100 mas for the
SuperCOSMOS data due to zonal systematic errors ranging from 50--200
mas when compared to 2MASS data.

To identify previously known high proper motion (HPM) stars in the
UCAC3, a source list was compiled using the VizieR on-line data tool,
along with targeted supplements from published literature.  In the
north we used the LSPM-North catalog \citep{2005AJ....130.1247L}
containing 61977 new and previously found stars having proper motions
greater than 0$\farcs$15 yr$^{-1}$.  For the south we utilized many
surveys, notably including the Revised NLTT Catalog
\citep{2003ApJ...582.1011S}, which produced 17730 stars with proper
motions greater than 0$\farcs$15 yr$^{-1}$, and the RECONS efforts
(SCR stars).  For a full list of catalogs used, see the UCAC3 README
file.  While this list is not comprehensive, this effort tagged
roughly 51000 known HPM stars in UCAC3 over the entire sky.  These
previously identified HPM stars were given a mean position (MPOS)
number greater than 140 million and do not have derived UCAC3 proper
motions. We instead used the proper motion data from the catalogs
themselves (see $\S$4.5).

Proper motion errors in the UCAC3 catalog for stars brighter than
R$\sim$12 are only $\sim$1--3 mas/yr in part because of the large
epoch spread of roughly 100 years in some cases.  The errors of the
fainter stars range from $\sim$2--3 mas/yr if found in SPM4 and
$\sim$6--8 mas/yr if SuperCOSMOS data are used in lieu of SPM4 data.

\subsection {SEARCH CRITERIA}

In this second paper we survey the southern sky between declinations
$-$47$\degr$ and 0$\degr$ using the same proper motion range as in
U3PM1, 0$\farcs$40 yr$^{-1}$ $>$ $\mu$ $\ge$ 0$\farcs$18 yr$^{-1}$.
In this area of the sky we identify an initial sample of 212356 proper
motion candidates.  We utilize the same search criteria as in U3PM1,
using UCAC3 flags with values indicative of real proper motion
objects.  A visual check from a sample of stars confirmed that these
flags still hold true in the region of the sky being surveyed.  All
stars must (1) be in the 2MASS catalog with an e2mpho (2MASS
photometry error) less than or equal to 0.05 magnitudes in all three
bands, (2) have a UCAC fit model magnitude between 7 and 17 mag, (3)
have a double star flag (dsf) equal to 0, 1, 5 or 6, meaning a single
star or fitted double, (4) have an object flag (objt) between $-$2 and
2 to exclude positions that used only overexposed images in the fit,
(5) have an MPOS number less than 140 million, to exclude already
known high proper motion stars, and (6) have a LEDA galaxy flag of
zero, meaning that the source is not in the LEDA galaxy catalog.
After all these cuts, there remain 17516 ``good'' candidate list, fewer
than expected for this region of the sky, when compared to 9248 in
U3PM1.  A total of 7641 candidates were excluded from the ``good''
candidates due to being marked as previously known in the UCAC3
catalog (MPOS number greater than 140 million).

These candidates were then cross-checked via VizieR and SIMBAD to
determine if they were previously known.  All cross-checks are
performed using a 90$\arcsec$ search radius, with one exception (the
NLTT catalog). A larger search radius of 180$\arcsec$ was used when
comparing UPM candidates to the NLTT and LHS catalog, which have been
found to have inaccurate positions as reported in
\citep{2002ApJS..141..187B}.  Thus, UCAC3 proper motion candidates
with positions differing from Luyten's or any other known object by
less than 90$\arcsec$ are considered known. Those differing from
Luyten's by 90--180$\arcsec$ are considered new discoveries but are
noted as possible NLTT stars in the tables. Those differing by
more than 180$\arcsec$ from Luyten are considered new discoveries.
All candidates matched to known stars had a final check to determine
if the proper motion and magnitudes matched --- those that match are
considered known and not reported in this sample.  As in U3PM1, it is
not a goal of this paper to revise the NLTT catalog and assign proper
identifications and accurate positions to NLTT entries; rather, the
goal is to identify new high proper motion stars.

After this, in effect, second cross-check for previously known stars,
the list was reduced to a manageable 3736 candidate proper motion
objects.  The 13780 known objects found during this cross-check shows
how incomplete the UCAC3 catalog can be in identifying previously
known high proper motion objects with the given search criteria.  Each
of these candidates was then visually inspected to confirm proper
motion by blinking the $B_J$ and $R_{59F}$ SuperCOSMOS digitized plate
images.  During blinking, we noticed that for declinations between
roughly $-$33$\degr$ and 0$\degr$ the epoch spread was insufficient
($\sim$3--5 years) to visually verify proper motion for all
candidates.  For those candidates, a second sweep was done by blinking
the $POSS-I R$ and $R_{59F}$ SuperCOSMOS digitized plate images.
Nearly 87\% of the candidates were found to have no verifiable proper
motions and were discarded.  The final counts of new discoveries are
500 proper motion objects in 474 systems.  Among these are 25 multiple
systems (24 doubles and one triple), of which ten were found to have
CPM to previously known primaries.

For this search we find a successful hit rate --- defined as the
number of new and known proper motion stars (21921) divided by the
total ``good'' candidates extracted (25157, including stars with an
MPOS number $>$ 140 million) --- of 87.1\%, which is higher than the
81.4\% hit rate found in TSN25.  After looking into the calculation
used in U3PM1 to determine the successful hit rate a counting error
was found.  The number of real objects excluded the known proper
motion objects tagged in the UCAC3 catalog (stars with an MPOS number
$>$ 140 million).  If we add these stars in the total for the U3PM1
count, we get a total of 7975 real objects giving a new successful hit
rate of 86.2\%, which is comparable to this paper.  At least three
factors mentioned in U3PM1 have been identified that can lead to false
detections in the UCAC3 proper motion survey.  First, some real
objects are discarded during the sifting mentioned above, particularly
because of the 2MASS criterion which states that $JHK_s$ photometry
errors must be less than 0.05 mag.  Second, the UCAC3 contains many
phantom proper motion objects due to incorrect matches during proper
motion calculations.  Third, other misidentifications arise from
blended images, where a single source in an earlier epoch catalog can
be matched with two stars in the UCAC3 data.

\section {RESULTS}

In Table~\ref{U3-discoveries}, we list the 474 new proper motion
stellar systems (500 objects) discovered during this search.  We
highlight the five red dwarf systems estimated to be within 25 pc
in Table~\ref{U3-distance}.  In both tables we list names,
coordinates, proper motions, 1$\sigma$ errors in the proper motions,
plate magnitudes from SuperCOSMOS, near-IR photometry from 2MASS, the
computed $R_{59F}-J$ color, a distance estimate, and notes.

\subsection{Positions and Proper Motions}

All positions on the ICRF system, proper motions, and errors are taken
directly from UCAC3, unless otherwise noted.  For a few stars that
were found during visual inspection without any UCAC3 data,
information has been obtained from alternate sources (see $\S$3.4).  For
this sample, the average positional errors reported in the UCAC3
catalog are 51 mas in RA and 50 mas in Dec.  For proper motions, the
average errors reported in the UCAC3 for this sample are 8.0 mas/yr in
$\mu_{\alpha}\cos\delta$ and 7.7 mas/yr in $\mu_{\delta}$.

\subsection{Photometry}

In Tables~\ref{U3-discoveries} and \ref{U3-distance}, we give
photographic magnitudes from the SuperCOSMOS and 2MASS surveys.  From
SuperCOSMOS, magnitudes are given from three plate emulsions, $B_J$,
$R_{59F}$, and $I_{IVN}$.  Magnitude errors are typically less than
0.3 mag for stars fainter than $\sim$15, with errors increasing for
brighter sources.  From 2MASS, $JHK_s$ infrared photometry is given,
with errors typically 0.05 mag or less due to the search criteria.
Additional objects found during visual inspection are typically
fainter with larger photometric errors.  The $R_{59F}-J$ color has
been computed to indicate the star's color.  

While SuperCOSMOS magnitudes are reported in the UCAC3, this sample
was checked against the SuperCOSMOS catalog to rectify some mismatches
found in the UCAC3 catalog.  In some cases, SuperCOSMOS magnitudes are
not given in the tables, due to blending, no source detection, high
chi-square or other problems where no reliable magnitude is available.
2MASS magnitudes are given for all but one object which was found
visually that is not present in the 2MASS catalog, as indicated in the
notes.

\subsection{Distances}

Plate photometric distance estimates are computed using the same
method as in U3PM1 and previous SCR searches.  Using the relations in
\citep{2004AJ....128..437H}, 11 distance estimates are generated based
on colors computed from the six-band photometry.  This method assumes
all objects are main sequence stars, and provides distances accurate
to 26\%, determined from the mean differences between the true
distances for stars with accurate (errors less than 10 mas)
trigonometric parallaxes and distances estimated from the relations.
Errors are higher for stars with missing photometry, resulting in
fewer than 11 relations, and stars that are not single, main sequence
red dwarfs, e.g.~cool subdwarfs and white dwarfs.  It is possible to
produce a distance with only one relation; however, six are needed to
be considered ``reliable'' because that allows for one magnitude
dropout.  Stars having fewer than six relations are identified in the
notes to Tables 2 and 3.  If a star is identified as a possible
subdwarf, the distance estimate is expected to be too large and is
given in brackets.

\subsection{Additional Objects}

In Table~\ref{U3-discoveries} we include 17 additional proper motion
objects found during visual inspection of the candidate fields.  These
objects are CPM companion candidates that either have fainter limiting
magnitudes than implemented for this search, were eliminated from the
candidate list by the search criteria, or have UCAC3 proper motions
less than 0$\farcs$18 yr$^{-1}$.  These new visual discoveries have
all been cross-checked with VizieR and SIMBAD using the same methods
described above for the main search.  Proper motions have been
obtained from UCAC3, SPM4, PPMXL \citep{2010AJ....139.2440R}, or
SuperCOSMOS, in that order.  For stars that were not found in the
UCAC3 data, positions were computed using the epoch, coordinates, and
proper motion obtained from the corresponding catalog.  Magnitudes are
obtained using the 2MASS and SuperCOSMOS catalogs to compute distance
estimates.

\section {ANALYSIS}

\subsection {Color-Magnitude Diagram}

\begin{figure}
 \epsscale{1.00}
 \includegraphics[angle=-90,scale=0.40]{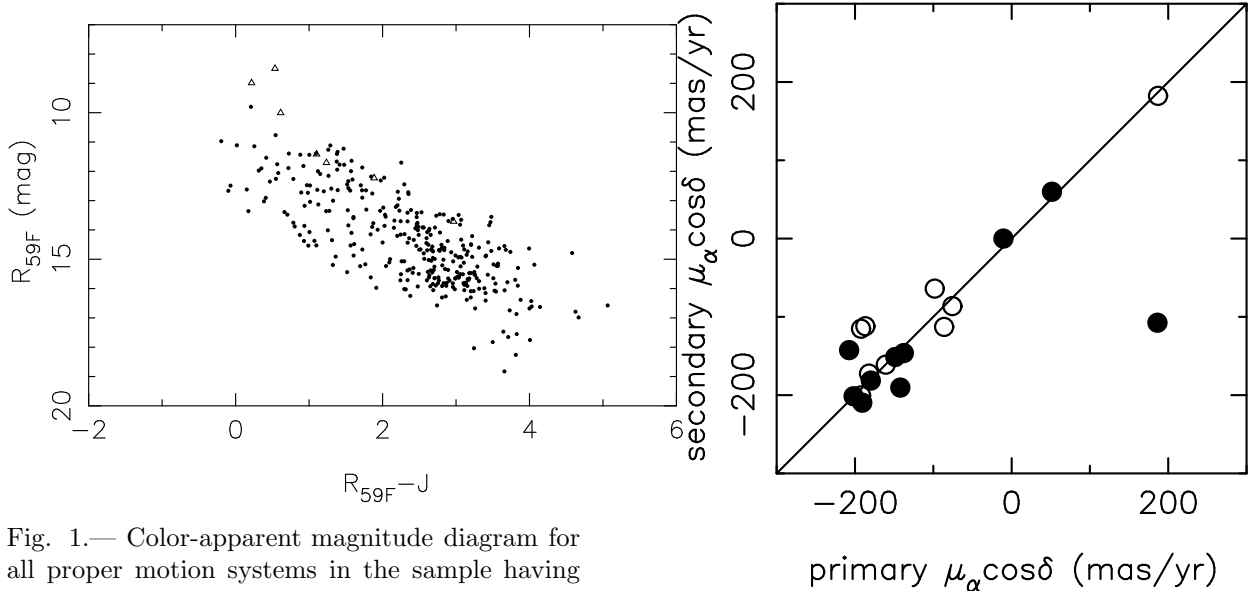}
 \caption{Color-apparent magnitude diagram for all proper motion
 systems in the sample having an $R_{59F} - J$ color.  New proper
 motion objects are represented by solid circles while known objects
 (CPM companions to new objects) are represented with open triangles.
 Data below $R_{59F} =$ 17 are CPM candidates noticed during
 visual inspection.}\label{color}
 \end{figure}

In Figure~\ref{color} we show a color-magnitude diagram of the 334 new
UPM proper motion objects (solid circles) and seven known objects (open
triangles, companions to UPM objects) from this search having $R_{59F}-J$
colors.  Symbols that fall below $R_{59F} \sim 17$ are CPM companion
candidates noticed during visual inspection.  The brightest new
object, UPM 0747-2537A, has $R_{59F}$ = 9.80 and is estimated to be at
a distance of 40.6 pc.  The reddest object found in this search is UPM
1848-0252 with $R_{59F}-J$ = 5.06, $R_{59F}$ = 16.57, at an estimated
distance of 26.9 pc.

The subdwarf population is not as well defined as in TSN18 and TSN25
because there are far fewer new objects.  Nonetheless, a separation
can be seen below the concentration of main sequence stars.  

\subsection {Reduced Proper Motion Diagram}

\begin{figure}
 \epsscale{1.00}
 \includegraphics[angle=-90,scale=0.40]{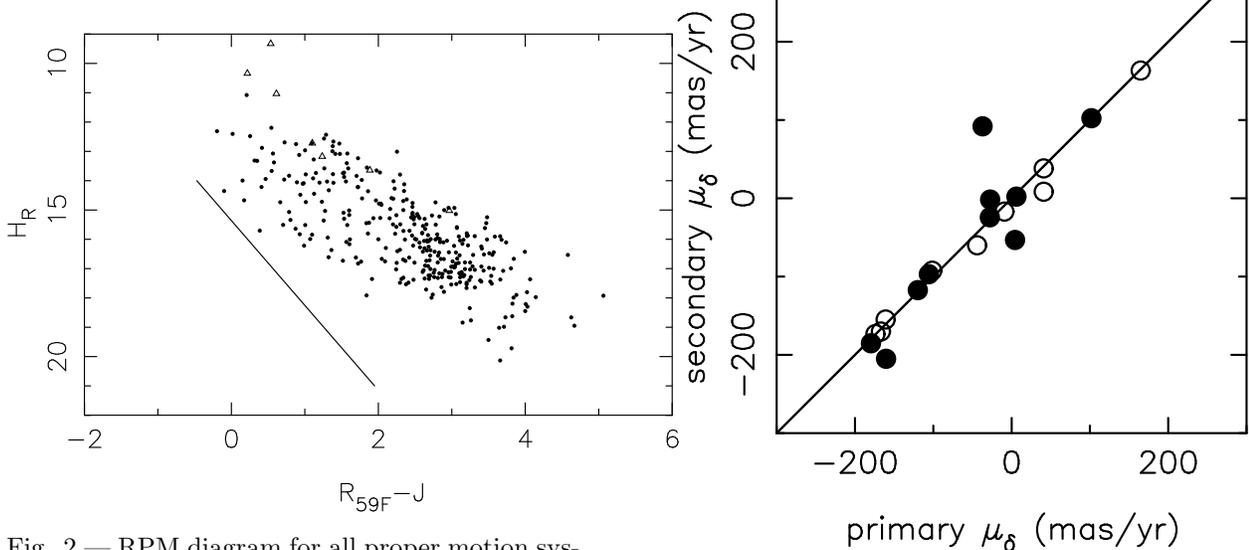}
 \caption{RPM diagram for all proper motion systems in this sample
 having an $R_{59F} - J$ color.  New proper motion objects are
 represented by solid circles while known objects (CPM companions to
 new objects) are represented with open triangles.  The empirical line
 separates the subdwarfs from where white dwarf candidates would be found.
 No white dwarf candidates were found in the current search.}\label{rpm}
 \end{figure}

In Figure~\ref{rpm}, we show the reduced proper motion (RPM) diagram
for all objects also plotted in Figure~\ref{color}, with similar
symbols for new and known objects.  The RPM diagram is a good method to
help separate white dwarfs and subdwarfs from main-sequence stars,
under the assumption that objects with larger distances tend to have
smaller proper motions.  Using the same method as in U3PM1 and TSN25
we obtain $H_{R_{59F}}$ via a modified distance modulus equation, in
which $\mu$ is substituted for distance:

\begin{displaymath}
H_{R_{59F}} = R_{59F} + 5 + 5\log\mu.
\end{displaymath}

The solid line seen in Figure~\ref{rpm} is used to separate white
dwarfs from subdwarfs.  This is the same empirical line used in U3PM1
and previous TSN papers.  No white dwarf candidates have been found
during this latest search.

Subdwarf candidates have been selected using the same method as in
U3PM1 and TSN25 --- stars with $R_{59F} - J >$ 1.0 and within 4.0 mag
in $H_{R}$ of the empirical line separating the white dwarfs are
considered subdwarfs.  From this survey there are 17 subdwarf
candidates, all with distance estimates greater than 122 pc, with the
exception of one, UPM 1712-4432, with an estimated distance of 33.9
(see $\S$4.4).  Because the relations used to estimate distances
assume that stars are on the main sequence, underluminous cool
subdwarfs and white dwarfs have large distances, which can, in fact,
be used to identify such objects.  The distance estimates for these
stars are presumably erroneous and are given in brackets in
Tables~\ref{U3-discoveries}, \ref{U3-distance} and \ref{U3CPM}.
Follow-up spectroscopic observations will be needed to confirm all
subdwarf candidates.

\subsection {New Common Proper Motion Systems}

In this search, we find 25 CPM candidate systems consisting of 24
binaries and one triple.  Included in these CPM systems are 16 new
systems and nine known systems with newly discovered components.

One binary system, UPM 0800-0617AB is a possible subdwarf binary
system.  The lone triple is an SCR system with two newly discovered
components.  In Table~\ref{U3CPM}, we list the CPM system primaries
and companions, their proper motions, and the companions' separations
and position angles relative to the primaries (defined to be the
brightest star in each system using the UCAC bandpass, or an alternate
bandpass if a UCAC value is not available).  We also provide distance
estimates for each component, where possible.  Components were
determined to be potentially physically associated using distance
estimates in conjunction with the proper motions and visual
inspections.  However, most of the companions were found during visual
inspection, meaning that proper motions, 2MASS and/or SuperCOSMOS
magnitudes may be missing or suspect, as identified in the notes.  For
systems with data missing in Table~\ref{U3CPM}, the physical
connection of the system components should be considered tentative.

\begin{figure}
 \epsscale{1.00}  
 \plotone{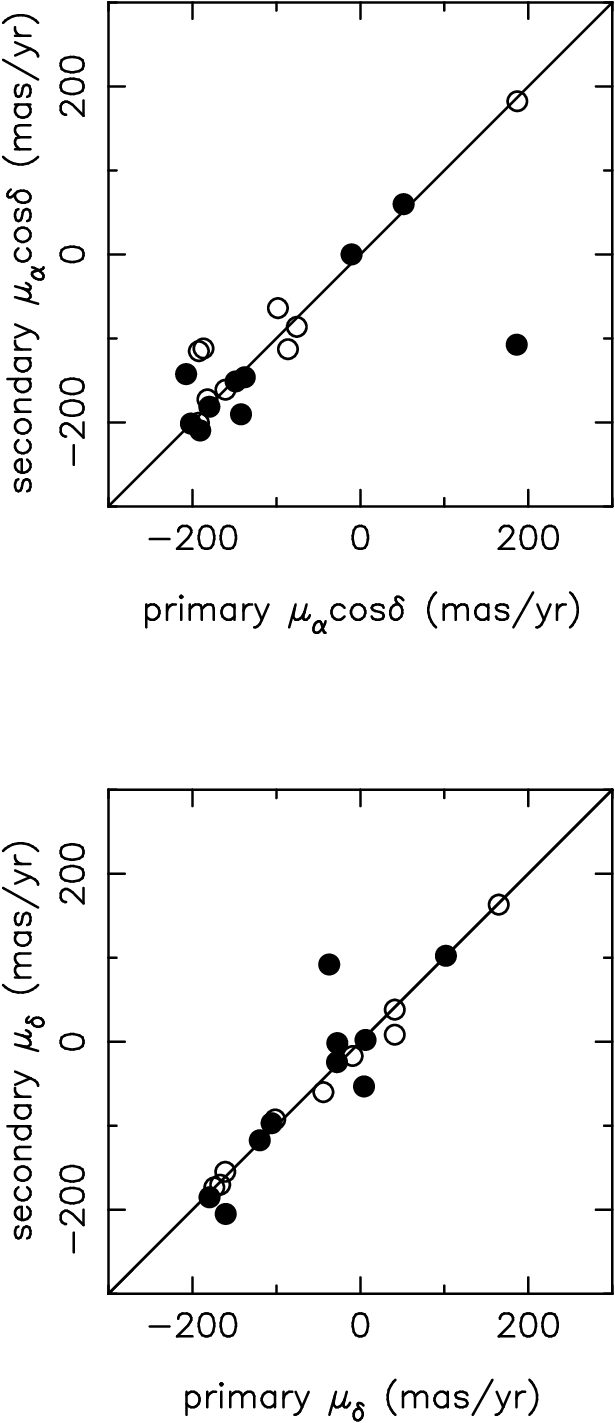} 
 \caption{Comparisons of proper motions in each coordinate,
   $\mu_{\alpha}\cos\delta$ (top) and $\mu_{\delta}$ (bottom), for
   components in CPM systems. Proper motions from the UCAC3 catalog
   are represented by solid circles while proper motions manually
   obtained through other means are denoted by open circles.  The
   solid line indicates perfect agreement.  Information on the
   outliers can be found in $\S$4.4}\label{cpm1}
 \end{figure}

In Figure~\ref{cpm1}, we show comparisons of the proper motions in
each coordinate for the 19 CPM systems for which both components have
a listed proper motion.  CPM candidates having proper motions from the
UCAC3 are represented by solid circles while those with proper motions
from other sources are represented by open circles.  If a proper motion
was not present in the UCAC3, data were obtained manually from the
SPM4, PPMXL or SuperCOSMOS databases, in that order.

\subsection{Notes on Specific Stars}

{\bf UPM 0443-4129AB} is a possible CPM binary.  However, UPM
0443-4129A has a suspect proper motion and the companion's distance
estimate uses fewer than 6 relations.  It is possible that this pair
is a case of a chance alignment.  See Table~\ref{U3CPM} for more
details.

{\bf BD-04 2807AB} is a possible CPM binary.  However, the primary has
a suspect proper motion, a distance estimate that uses fewer than 6
relations, and there is no distance estimate for the secondary.  It is
possible that this pair is a case of a chance alignment.  See
Table~\ref{U3CPM} for more details.

{\bf UPM 0747-2537A} is the brightest new discovery from this search
with $R_{59F} =$ 9.80 and an estimated distance of 40.6 pc.  However,
only one relation was viable, making the distance estimate unreliable.

{\bf UPM 0800-0617AB} is a possible candidate for a binary subdwarf
system.  The primary is a possible subdwarf at an estimated distance
of 175.5 pc.  The secondary is at a separation of 5.8$\arcsec$ at
position angle 297.2$^{\circ}$ from the primary. Color information is
insufficient for a reliable distance estimate.

{\bf UPM 1226-3516B and C} are in a candidate triple system with SCR
1226-3515A.  The A and B components are separated by 49.8$\arcsec$ at
a position angle of 191.3$^{\circ}$.  The C component has a separation
of 97.0$\arcsec$ at a position angle of 146.9$^{\circ}$ from the
primary.  The C component has a suspect proper motion and the distance
estimates for all there components are inconsistent.  In particular,
the C component may not be a part of the system.  See
Table~\ref{U3CPM} for more details.

{\bf UPM 1712-4432} is a subdwarf candidate with $R_{59F} =$ 13.04 and
$R_{59F}-J =$ 1.01 at a distance of 33.9 pc.  However, only three
relations were viable, making the distance estimate unreliable.
SuperCOSMOS magnitudes are indicative of a blended image, meaning this
is likely not what it seems.

{\bf UPM 1718-2245B} has an estimated distance of only 13.2 pc based
on 7 relations, making it the nearest candidate in the sample.
However, the primary has a distance estimate of 25.4 pc based on 10
relations so we favor the larger distance for the system. 

{\bf UPM 1848-0252} is the reddest new discovery from this search,
with $R_{59F}-J =$ 5.06 and an estimated distance of 26.9 pc.

\subsection {Comparison to Previous Proper-Motion Surveys}

\begin{figure}
 \epsscale{1.00}  
 \plotone{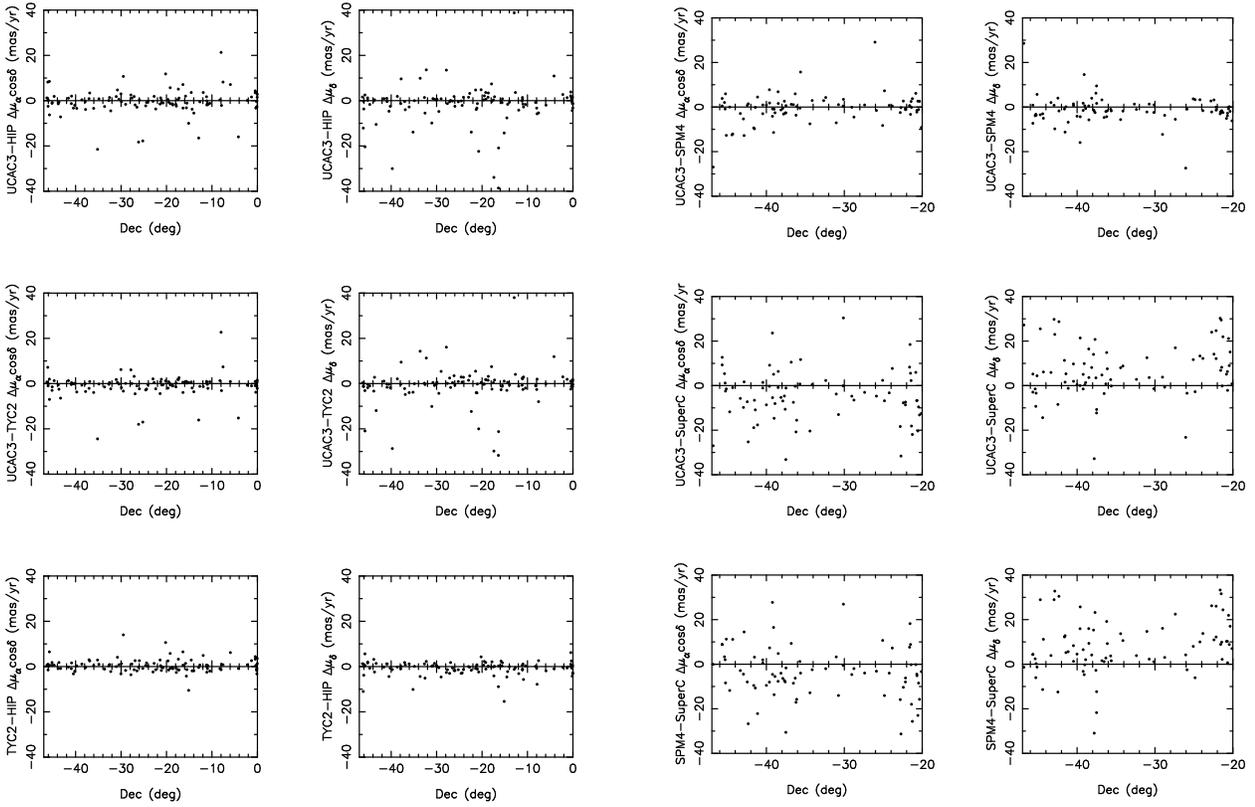} 
 \caption{Comparisons of UCAC3, Hipparcos and Tycho-2 proper motions per
   coordinate, $\Delta\mu_{\alpha}\cos\delta$ (left column) and
   $\Delta\mu_{\delta}$ (right column). }\label{pm1}
 \end{figure}

During production of the UCAC3 catalog, we made an effort to tag
previously known HPM stars.  For these stars, proper motions were taken
from their respective catalogs rather than calculated using UCAC3
methodology, which made comparisons to other catalogs/surveys
difficult.  However, during the present search we have found 104 stars
in both the Hipparcos and Tycho-2 catalogs that are not tagged as HPM
stars in the UCAC3 catalog --- these stars are proper motion
candidates that were found to be in Tycho-2 during cross-checking.  A
2.5$\arcsec$ radius was used to match these stars to sources in the
Hipparcos catalog so that we can compare the bright end of the UCAC3
proper motion stars ($R$ $\sim$ 7.13-13.66) to stars in both the
Tycho-2 and Hipparcos catalogs.  In Figure~\ref{pm1}, we compare
proper motions in RA and Dec for these stars as given in UCAC3,
Hipparcos, and Tycho-2.  These plots show that the differences in
proper motions are small, in general less than 10 mas/yr, and no
significant systematic errors as a function of declination are seen.
The RMS differences between UCAC3 proper motions in
$\Delta\mu_{\alpha}\cos\delta$ and $\Delta\mu_{\delta}$ and those from
Hipparcos are 5.7 and 9.1 mas/yr, respectively.  Comparisons to
Tycho-2 yield RMS differences of 5.2 and 8.3 mas/yr, respectively.
Lower RMS differences of 3.0 mas/yr in $\Delta\mu_{\alpha}\cos\delta$
and 3.2 mas/yr in $\Delta\mu_{\delta}$ are seen when comparing the
Hipparcos to Tycho-2 proper motions.

\begin{figure}
 \epsscale{1.00}  
 \plotone{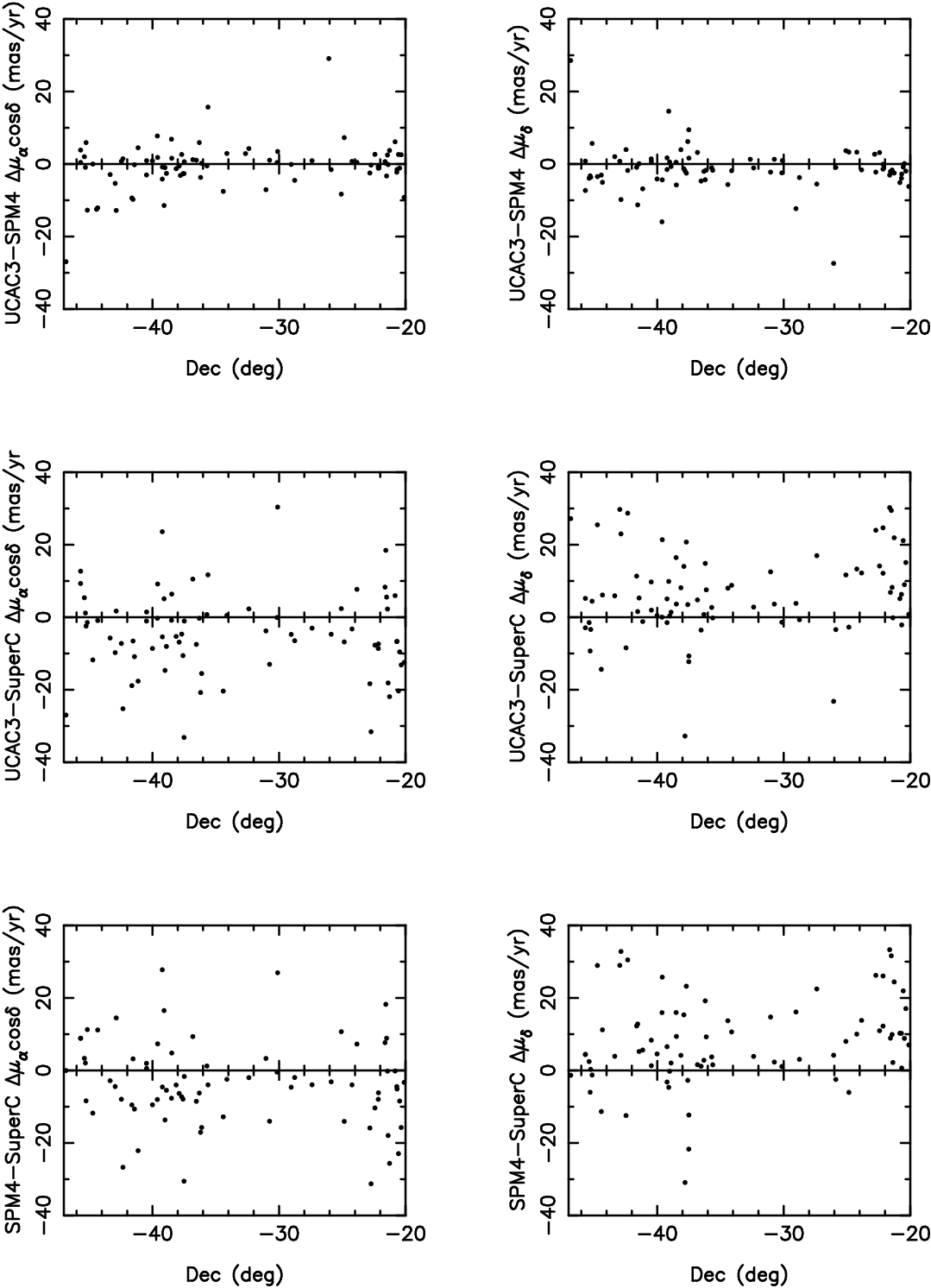} 
 \caption{Comparisons of UCAC3, SuperCOSMOS and SPM4 proper motions per
   coordinate, $\Delta\mu_{\alpha}\cos\delta$ (left column) and
   $\Delta\mu_{\delta}$ (right column). }\label{pm2}
 \end{figure}

To investigate the fainter end of UCAC3, we compare results for 77
stars ($R$ $\sim$ 10.88-16.69) that are in both the SPM4 and
SuperCOSMOS catalogs that were not tagged as HPM stars in the UCAC3
catalog --- these stars are proper motion candidates that were found
to be SCR stars during cross-checking.  A 2.5$\arcsec$ radius was used
to match these stars to sources in the SPM4 catalog.  The SPM4 catalog
only covers Dec = $-$90 to $-$20 sky area, limiting the area included for
this comparison.  In Figure~\ref{pm2}, we compare proper motions in RA
and Dec for these stars as given in UCAC3, SuperCOSMOS, and SPM4.
These plots show that differences in proper motions are similar to
those found for brighter stars when comparing UCAC3 and SPM4, but the
differences are much larger for the SuperCOSMOS results.  The RMS
differences between UCAC3 proper motions in
$\Delta\mu_{\alpha}\cos\delta$ and $\Delta\mu_{\delta}$ and those in
SPM4 are 6.0 and 5.7 mas/yr respectively.  Comparisons to SuperCOSMOS
yield RMS differences of 16.5 and 14.1 mas/yr, respectively.  In
Figure~\ref{pm2}, we also see that proper motions in Dec appear to be
systematically shifted in the SuperCOSMOS data.  These high RMS
results and the systematic shift are also seen in the comparison of
the SPM4 to the SuperCOSMOS proper motions, yielding RMS differences
of 15.6 and 15.2 mas/yr in $\Delta\mu_{\alpha}\cos\delta$ and
$\Delta\mu_{\delta}$, respectively.  The higher RMS differences for
the SuperCOSMOS proper motions are in agreement with the findings of
TSN18 and U3PM1 where SCR proper motions were found to have higher RMS
differences when compared to other external catalogs.  It is worth
noting that the SuperCOSMOS proper motion RMS reported here are not
representative of the entire catalog.  Objects having an R$\sim$16--19
with $\mu >$ 0$\farcs$10 yr$^{-1}$ in the SuperCOSMOS catalog should
have an RMS no greater than 10 mas/yr, and considerably better for
fields with decades between the epochs (See Tables 1 and 3 from
\citep{2001MNRAS.326.1315H}).

Random and systematic differences of order 10 mas/yr in proper motions
between the various catalogs, particularly at the faint end, are
expected because of different data quality, measurements, reductions
and epoch differences.  SuperCOSMOS for example uses Schmidt plates
for both early and recent epoch which typically show large errors.  The
proper motions of faint stars in UCAC3 are based on early epoch
Schmidt plates for the sky area north of $-$20 deg Dec and CCD
observations for recent epoch data.  A combination of CCD data and
early astrograph data (SPM plates) is used south of $-$20 deg, with
significantly smaller errors.  The SPM4 proper motions are derived
entirely on SPM astrograph plates from 2 epochs.  At the bright end
proper motions are more reliable due to higher quality of Hipparcos
and Tycho data as well as availability of many other star catalogs,
most of which have been used in common between Tycho-2 and UCAC3.
However, there can be large differences between Hipparcos and Tycho-2
for some stars because the Hipparcos PMs are based on only about 3.5
years of observing (although with high quality), while Tycho-2 PMs are
based on typically 100 years epoch difference.  Multiplicity and
residual orbital motions sometimes render Hipparcos PMs inferior in
spite of their small formal astrometric errors.

In TSN25 a total of 3073 objects were reported, all of which fit
within the proper motion and declination constraints of this paper.
During this UCAC3 search, only 770 of the 3073 objects reported in
TSN25 were recovered, or a low 25.1\% recovery rate.  This is
primarily due to the UCAC3 catalog having no proper motion or a
reported proper motion not meeting the criteria of this paper
(0$\farcs$40 yr$^{-1}$ $>$ $\mu$ $\ge$ 0$\farcs$18 yr$^{-1}$) for
$\sim$70\% of the new discoveries listed in TSN25.

The Hipparcos catalog contains 118218 total objects, of which 1690
meet the proper motion and declination constraints of this paper.
Tycho-2 contains 2539913 total objects in the main catalog, of which
3187 meet similar limits.  We recover 1316 Hipparcos stars and 2543
Tycho-2 stars using the search criteria of this paper, yielding
recovery rates of 77.9\% and 79.8\% respectively.  Objects missed in
this UCAC3 survey are primarily due to UCAC3 lacking a source
detection for $\sim$15\% of the Tycho-2 objects.  The relatively high
recovery rates of UCAC3, when compared to the Hipparcos and Tycho-2
catalogs, implies the UCAC3 can be used as a reliable source to search
for new proper motion stars with $\mu$ = 0.18--0.40$\arcsec$ yr$^{-1}$
for other portions of the sky.

\section {DISCUSSION}

We have completed a sweep of the southern sky for new proper motion
systems using the UCAC3 catalog.  So far, we have uncovered 916 new
proper motion systems, of which 474 are described in this paper.
These systems constitute an increase of 19.4\% over the total number of
SCR systems discovered in the southern sky and an increase of 20.7\%
over SCR systems in the southern sky with 0$\farcs$40 yr$^{-1}$ $>$
$\mu$ $\ge$ 0$\farcs$18 yr$^{-1}$.  This UCAC3 proper motion survey
has added 3.8\% to the list of entries in the NLTT catalog south of
Dec $=$ 0$\degr$ with 974 new proper motion objects from U3PM1 and
this paper.

\begin{figure}
 \epsscale{1.00}
 \includegraphics[angle=-90,scale=0.40]{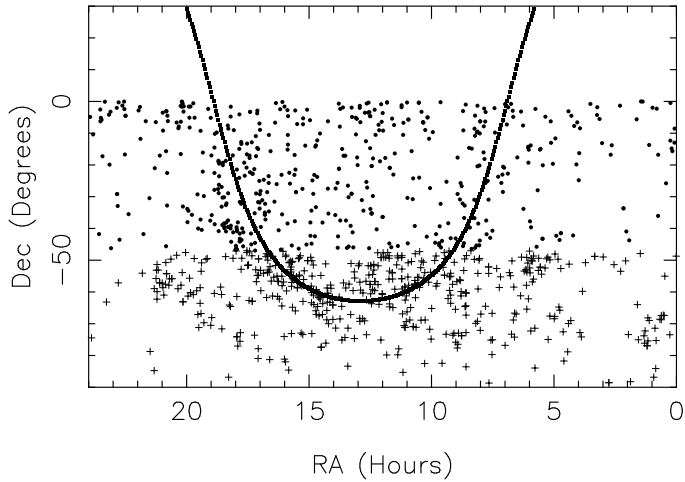}
 \caption{Sky distribution of all UCAC3 proper motion survey objects
 reported in U3PM1 (plus signs) and this paper (solid circles),
 i.e.~those between declinations $-$90$\degr$ and 0$\degr$ having
 0$\farcs$40 yr$^{-1}$ $>$ $\mu$ $\ge$ 0$\farcs$18 yr$^{-1}$.  The
 curve represents the Galactic plane. }\label{sky}
 \end{figure}

In Figure~\ref{sky}, we show the sky distribution of systems found to
date during the UCAC3 proper motion survey.  Plus signs represent
objects from U3PM1 and solid circles represent objects described in
this paper.  Overall, the distribution of new objects is similar to
that seen in Figure 6 of TSN25, including the discovery of many new
proper motion systems along the Galactic plane.

\begin{figure}
 \epsscale{1.00}  
 \includegraphics[angle=-90,scale=0.40]{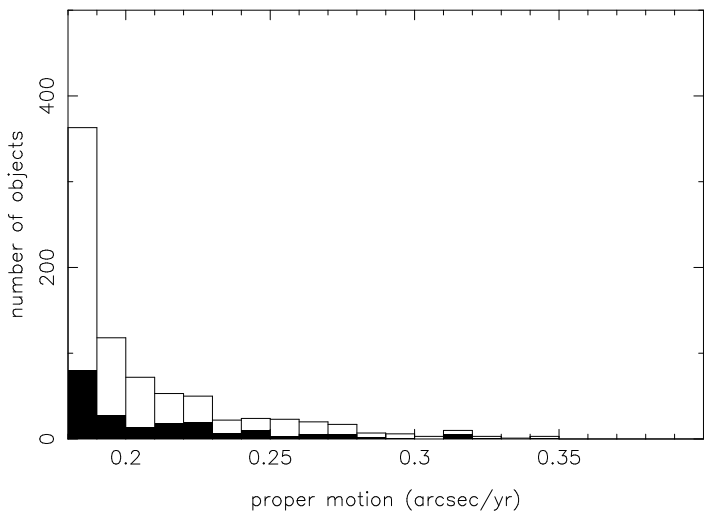}
 \caption{Histogram showing the number of proper motion objects in
 0$\farcs$01 yr$^{-1}$ bins for the entire UCAC3 proper motion sample
 (empty bars) and the number of those objects having distance
 estimates within 50 pc (filled bars).  }\label{hist}
 \end{figure}

In Figure~\ref{hist}, we show a histogram of the number of proper
motion systems discovered to date during the UCAC3 proper motion
survey, in 0$\farcs$01 yr$^{-1}$ bins, and highlighting the number of
those having distance estimates within 50 pc.  Predictably, this plot
shows that the slowest proper motion bins have the most new systems.
This confirms the trend reported in TSN18, TSN25 and U3PM1, and
suggests once again that more nearby stars are yet to be found at
slower proper motions.

We have found a total of 57 CPM candidate systems during this UCAC3
proper motion survey, including 55 binaries and two triples.  These
systems have separations of 1--359$\arcsec$ and will need further
investigation to confirm which of the systems are, in fact,
gravitationally linked.  In addition, we have revealed a total of 48
subdwarf candidates, each of which is worthy of followup observations,
given the scarcity of nearby subdwarfs.  Finally, we have found 20 
red dwarf systems likely to be within 25 pc.  We plan to obtain CCD
photometry through $VRI$ filters for stars having estimated distances
within 25 pc in order to make more reliable distance estimates using
the $VRIJHK$ relations presented in \citep{2004AJ....128.2460H}.
Stars estimated to be within 10 pc will then be put on the CTIO
parallax program, potentially to join the ranks of the few hundred
systems known to be so close to the Sun \citep{2006AJ....132.2360H}.
 

\acknowledgments

We thank the entire UCAC team for making this proper motion survey
possible, and the USNO summer students, who helped with tagging HPM
stars in the UCAC3 catalog.  Special thanks go to members of the
RECONS team at Georgia State University for their support, and John
Subasavage in particular for assistance with the SCR searches.  This
work has made use of the SIMBAD, VizieR, and Aladin databases operated
at the CDS in Strasbourg, France.  We have also made use of data from
the Two-Micron All Sky Survey, SuperCOSMOS Science Archive and the
Southern Proper Motion catalog.


\clearpage


 \clearpage

 \clearpage
 
\clearpage



\begin{deluxetable}{lrrl}
\tabletypesize{}

\tablecaption{New Proper Motion Systems from the UCAC3 and SCR proper
  motion surveys\label{diststat}}

\tablewidth{0pt}
\tablehead{
\colhead{Paper}             &
\colhead{New Systems}       &
\colhead{New Systems}       &
\colhead{References}       \\

\colhead{}                  &
\colhead{total}             &
\colhead{$\le$ 25 pc}       &
\colhead{}                  }

\startdata
  U3PM1      &  442 &  15 & \citep{upm1}                \\
  U3PM2      &  474 &   4 & this paper                  \\
             &      &     &                             \\
  TSN08      &    5 &   2 & \citep{2004AJ....128..437H} \\
  TSN10      &    4 &   4 & \citep{2004AJ....128.2460H} \\
  TSN12      &  141 &  12 & \citep{2005AJ....129..413S} \\
  TSN15      &  152 &  25 & \citep{2005AJ....130.1658S} \\
  TSN18      & 1605 &  30 & \citep{2007AJ....133.2898F} \\
  TSN25      & 2817 &  79 & \citep{tsnxxv}              \\
             &      &     &                             \\
  totals     & 5640 & 171 &                             \\
\enddata
\end{deluxetable}
\clearpage

\begin{deluxetable}{lrrrrrrrrrrrrrrl}
\tabletypesize{\tiny} 
\rotate

\tablecaption{New UCAC3 High Proper Motion Systems between
  Declinations $-$47$\degr$ and 0$\degr$ with 0$\farcs$40
  yr$^{-1}$ $>$ $\mu$ $\ge$ 0$\farcs$18
  yr$^{-1}$ \label{U3-discoveries}} \tablewidth{0pt}

\tablehead{
\colhead{Name}                         &
\colhead{RA J2000.0}                   &
\colhead{DEC J2000.0}                  &
\colhead{$\mu_{\alpha}\cos\delta$}     &
\colhead{$\mu_{\delta}$}               &
\colhead{sig$\mu_{\alpha}$}            &
\colhead{sig$\mu_{\delta}$}            &
\colhead{$B_J$}                        &
\colhead{$R_{59F}$}                    &
\colhead{$I_{IVN}$}                    &
\colhead{$J$}                          &
\colhead{$H$}                          &
\colhead{$K_s$}                        &
\colhead{$R_{59F}$ $-$ $J$}            &
\colhead{Est Dist}                     &
\colhead{Notes}                        \\
			               
\colhead{}                             &
\colhead{(deg)}                        &
\colhead{(deg)}                        &
\colhead{(mas/yr)}                     &
\colhead{(mas/yr)}                     &
\colhead{(mas/yr)}                     &
\colhead{(mas/yr)}                     &
\colhead{}                             &
\colhead{}                             &
\colhead{}                             &
\colhead{}                             &
\colhead{}                             &
\colhead{}                             &
\colhead{}                             &
\colhead{(pc)}                         &
\colhead{}                             }

\startdata

 UPM 0004-0833  &    1.1472383 &  -8.5664958 &   182.2 &   -26.3 &     8.6 &     8.6 &  14.680 &  12.649 &  11.643 &  10.874 &  10.255 &  10.072 &   1.775 &    63.8  &                                                                                                       \\
 UPM 0004-1258  &    1.2061531 & -12.9791222 &   102.0 &  -156.8 &    10.2 &     6.6 &  14.504 &  12.478 &  11.385 &  10.774 &  10.160 &   9.973 &   1.704 &    62.3  &                  											  \\
 UPM 0009-1539  &    2.4914975 & -15.6590597 &   184.0 &   -15.7 &     8.7 &     8.7 &  17.100 &  15.050 &  13.144 &  12.387 &  11.793 &  11.562 &   2.663 &    93.0  &                  											  \\
 UPM 0011-1448  &    2.9555050 & -14.8079581 &   179.8 &   -34.9 &     9.0 &     9.0 &  14.785 &  13.131 &  12.454 &  12.005 &  11.353 &  11.246 &   1.126 &   115.6  &                  											  \\
 UPM 0014-0029  &    3.6589169 &  -0.4939803 &   161.0 &  -104.7 &    13.7 &    14.0 &  17.819 &  15.829 &  13.947 &  12.292 &  11.708 &  11.457 &   3.537 &    58.7  &                  											  \\
 UPM 0014-1219  &    3.7410256 & -12.3317842 &   183.9 &   -21.6 &     7.0 &     3.9 &  17.189 &  15.068 &  13.388 &  12.808 &  12.230 &  11.974 &   2.260 &   130.3  &                  											  \\
 UPM 0025-2547  &    6.3606617 & -25.7849942 &   170.4 &    65.2 &     9.4 &     9.3 &  20.968 &  18.826 &  16.962 &  15.167 &  14.514 &  14.163 &   3.659 &   179.8  &   \tablenotemark{a}     										  \\
 UPM 0044-1647  &   11.1954375 & -16.7984897 &   187.8 &    29.7 &    14.6 &    14.1 & \nodata & \nodata &  13.280 &  12.405 &  11.777 &  11.549 & \nodata &    91.5  &   \tablenotemark{b}              									  \\
\enddata

\tablenotetext{a}{Proper motions suspect}
\tablenotetext{b}{Number of relations used for distance estimate $<$ 6: plate distance less reliable}
\tablenotetext{c}{SuperCOSMOS plate magnitudes suspect}
\tablenotetext{d}{Common proper motion companion; see Table~\ref{U3CPM}}
\tablenotetext{e}{Not detected during automated search but noticed by eye during the blinking process}
\tablenotetext{f}{Subdwarf candidate selected from RPM diagram; plate distance [in bracket] is incorrect}
\tablenotetext{g}{Possible NLTT star with a position difference $>$ 90$\arcsec$ when compared to UCAC3 position}

\end{deluxetable}


\begin{deluxetable}{lrrrrrrrrrrrrrrl}
\tabletypesize{\tiny} 
\rotate 

\tablecaption{New UCAC3 High Proper Motion Systems estimated to be
  within 25 pc between Declinations $-$47$\degr$ and 0$\degr$ with
  0$\farcs$40 yr$^{-1}$ $>$ $\mu$ $\ge$ 0$\farcs$18
  yr$^{-1}$ \label{U3-distance}} \tablewidth{0pt}

\tablehead{
\colhead{Name}                          &
\colhead{RA J2000.0}                    &
\colhead{DEC J2000.0}                   &
\colhead{$\mu_{\alpha}\cos\delta$}      &
\colhead{$\mu_{\delta}$}                &
\colhead{sig $\mu_{\alpha}$}            &
\colhead{sig $\mu_{\delta}$}            &
\colhead{$B_J$}                         &
\colhead{$R_{59F}$}                     &
\colhead{$I_{IVN}$}                     &
\colhead{$J$}                           &
\colhead{$H$}                           &
\colhead{$K_s$}                         &
\colhead{$R_{59F}$ $-$ $J$}             &
\colhead{Est Dist}                      &
\colhead{Notes}                         \\
			                
\colhead{}                              &
\colhead{(deg)}                         &
\colhead{(deg)}                         &
\colhead{(mas/yr)}                      &
\colhead{(mas/yr)}                      &
\colhead{(mas/yr)}                      &
\colhead{(mas/yr)}                      &
\colhead{}                              &
\colhead{}                              &
\colhead{}                              &
\colhead{}                              &
\colhead{}                              &
\colhead{}                              &
\colhead{}                              &
\colhead{(pc)}                          &
\colhead{}                              }

\startdata

 UPM 1349-4228  & 207.2552625 & -42.4784189 &  -161.9 &   -84.6 &     1.3 &     3.0 &  14.468 &  11.703 & \nodata &   9.449 &   8.863 &   8.622 &   2.254 &    24.4  &                                                                                                       \\
 UPM 1648-3459  & 252.1200667 & -34.9967942 &   178.6 &   142.5 &     4.4 &     4.4 & \nodata &  14.631 & \nodata &  10.687 &  10.161 &   9.907 &   3.944 &    22.1  &                                                                                                       \\
 UPM 1654-3105  & 253.6846164 & -31.0961000 &   -32.4 &  -215.9 &     7.4 &     7.2 &  15.122 &  13.553 &  11.977 &  10.072 &   9.482 &   9.237 &   3.481 &    23.8  &   \tablenotemark{a}                                                                                   \\
 UPM 1718-2245A & 259.6129147 & -22.7616683 &  -160.7 &  -160.8 &     7.2 &     6.8 &  15.469 &  13.836 &  13.155 &  10.385 &   9.806 &   9.572 &   3.451 &    25.4  &   \tablenotemark{b}                                                                                   \\
 UPM 1718-2245B & 259.6213031 & -22.7746183 &  -161.1 &  -154.8 &    10.9 &     9.6 & \nodata &  14.787 &  13.289 &  10.207 &   9.608 &   9.375 &   4.580 &    13.2  &   \tablenotemark{b}  \tablenotemark{c}                                                                \\
 UPM 1840-1934  & 280.1299600 & -19.5831464 &   201.2 &   -89.8 &     8.8 &     8.1 & \nodata &  13.733 &  12.400 &  10.264 &   9.706 &   9.450 &   3.469 &    20.7  &                                                                                                       \\
\enddata

\tablenotetext{a}{Possible NLTT star with a position difference $>$ 90$\arcsec$ when compared to UCAC3 position}
\tablenotetext{b}{Common proper motion companion; see Table~\ref{U3CPM}}
\tablenotetext{c}{Not detected during automated search but noticed by eye during the blinking process}

\end{deluxetable}


\begin{deluxetable}{lccclcccccl}
\rotate 
\tabletypesize{\scriptsize} 
\tablecaption{Common Proper Motion Candidate
Systems\label{U3CPM}}
\tablewidth{0pt}

\tablehead{
\colhead{Primary}                        &
\colhead{$\mu_{\alpha}\cos\delta$}       &
\colhead{$\mu_{\delta}$}                 &
\colhead{Distance}                       &
\colhead{Secondary/Tertiary}             &
\colhead{$\mu_{\alpha}\cos\delta$}       &
\colhead{$\mu_{\delta}$}                 &
\colhead{Distance}                       &
\colhead{Separation}                     &
\colhead{$\theta$}                       &
\colhead{notes}                          \\

\colhead{}                               &
\colhead{(mas/yr)}                       &
\colhead{(mas/yr)}                       &
\colhead{(pc)}                           &
\colhead{}                               &
\colhead{(mas/yr)}                       &
\colhead{(mas/yr)}                       &
\colhead{(pc)}                           &
\colhead{($\arcsec$)}                    &
\colhead{($\degr$)}                      &
\colhead{}                               }

\startdata

 UPM 0209-3339A          &  -86.1  & -166.9  &  49.5   & UPM 0209-3339B          & -112.9   &  -170.2  & \nodata &  11.6 &  78.1 & \tablenotemark{a}  \tablenotemark{b}                                                        \\
 UPM 0443-4129A          &  186.1  &    4.3  &  39.3   & 2MASS J04430760-4128575B& -107.5   &   -53.1  & \nodata &   6.8 & 339.2 & \tablenotemark{a}  \tablenotemark{b}  \tablenotemark{c}                                     \\
 UPM 0528-4313A          &  -75.6  &  164.7  &  70.4   & UPM 0528-4313B          &  -86.3   &   163.2  & 109.1   &  42.1 & 209.0 & \tablenotemark{a}                                                                           \\
 UPM 0659-0052A          &  -58.3  & -184.1  &  78.2   & UPM 0659-0052B          & \nodata  &  \nodata & \nodata &  13.8 & 151.6 & \tablenotemark{a}  \tablenotemark{b}  \tablenotemark{c}                                     \\
 UPM 0704-0602A          & \nodata & \nodata & 123.4   & UPM 0704-0602B          &   99.5   &  -153.0  &  37.8   &  12.2 & 359.1 & \tablenotemark{a}  \tablenotemark{b}  \tablenotemark{c}  \tablenotemark{d}                  \\
 UPM 0747-2537A          & -148.5  &  101.9  &  40.6   & UPM 0747-2537B          & -151.3   &   102.3  &  47.3   &  12.0 & 237.4 & \tablenotemark{b}                                                                           \\
 UPM 0800-0617A          &  135.2  & -233.8  &[175.5]  & UPM 0800-0617B          & \nodata  &  \nodata & \nodata &   5.8 & 297.2 & \tablenotemark{a}  \tablenotemark{b}  \tablenotemark{c}  \tablenotemark{e}                  \\
 BD-04  2807A            & -142.1  &  -37.1  &  19.5   & UPM 1009-0501B          & -190.5   &    92.1  & \nodata &  20.9 & 338.5 & \tablenotemark{a}  \tablenotemark{b}                                                        \\
\enddata

\tablenotetext{a}{Not detected during automated search but noticed by eye during the blinking process}
\tablenotetext{b}{Number of relations used for distance estimate $<$ 6; plate distance less reliable}
\tablenotetext{c}{Proper motions suspect}
\tablenotetext{d}{SuperCOSMOS plate magnitudes suspect}
\tablenotetext{e}{Subdwarf candidate selected from RPM diagram; plate distance [in bracket] is incorrect}
\tablenotetext{f}{Source not in 2MASS}
\tablenotetext{g}{Possible NLTT star with a position difference $>$ 90$\arcsec$ when compared to UCAC3 position}

\end{deluxetable}


\end{document}